\documentclass[12pt,preprint]{aastex}
\usepackage{graphicx}
\usepackage{epstopdf}
\usepackage[colorlinks,linkcolor=blue]{hyperref}

\shorttitle{Discovery of $\gamma$-ray emission of III Zw 2}

\shortauthors{Liao et al.}

\begin{document}

\title{Discovery of $\gamma$-ray emission from the radio-intermediate quasar III Zw 2: violent jet activity with intraday $\gamma$-ray variability}

\author{Neng-Hui Liao\altaffilmark{1},Yu-Liang Xin\altaffilmark{1,2}, Xu-Liang Fan\altaffilmark{2,3}, Shan-Shan Weng\altaffilmark{4}, Shao-Kun Li\altaffilmark{3},  Liang Chen\altaffilmark{5}, Yi-Zhong Fan\altaffilmark{1}}
\email{E-mail:liaonh@pmo.ac.cn (NHL); yzfan@pmo.ac.cn (YZF)}

\altaffiltext{1}{Key Laboratory of Dark Matter and Space Astronomy, Purple Mountain Observatory, Chinese Academy of Sciences, Nanjing 210008, China}

\altaffiltext{2}{Graduate University of Chinese Academy of Sciences, Yuquan Road 19, Beijing 100049, China}


\altaffiltext{3} {Key Laboratory for the Structure and Evolution of Celestial Objects, Yunnan Observatories, Chinese Academy of Sciences, Kunming 650011, China}

\altaffiltext{4} {Department of Physics, Nanjing Normal University, Nanjing, China}

\altaffiltext{5} {Key Laboratory for Research in Galaxies and Cosmology, Shanghai Astronomical Observatory, Chinese Academy of Sciences, 80 Nandan Road, Shanghai 200030, China}
\begin{abstract}
III Zw 2 is the prototype of radio-intermediate quasars. Although there is the evidence of possessing strong jet, significant $\gamma$-ray emission has not been reported before. In this work, we carry out a detailed analysis of the latest {\it Fermi}-LAT {\it Pass} 8 data. No significant $\gamma$-ray signal has been detected in the time-averaged 7-year {\it Fermi}-LAT data of III Zw 2, we however have identified two distinct $\gamma$-ray flares with isotropic luminosities of $\sim 10^{45}$ erg $\rm s^{-1}$. Multiwavelength data analysis (including also the optical photometric observations from Yunnan Observatories) are presented and the main finding is the simultaneous optical and $\gamma$-ray flares of III Zw 2 appearing in Nov. 2009. Violent $\gamma$-ray variability with doubling timescale of 2.5 hours was detected in another $\gamma$-ray flare in May 2010, for which the 3-hour $\gamma$-ray peak flux is $\sim 250$ times of the average flux in 7 years. Rather similar behaviors are observed in blazars and the blazar model can reasonably reproduce the spectral energy distribution of III Zw 2 in a wide energy range, strongly suggesting that its central engine resembles that of blazars. In view of its core which shares radio similarities with young radio sources together with the weak extended radio lobe emission, we suggest that III Zw 2 harbors a recurrent activity core and serves as a valuable target for investigating the fuelling and triggering of the activity in radio loud AGNs.
\end{abstract}


\section{INTRODUCTION}
Active galactic nuclei (AGNs), the most luminous and persistent sources of electromagnetic radiation in the universe, are powered by accretion of material onto super-massive black holes (SMBHs; Antonucci 1993; Urry \& Padovani 1995). In optically selected samples (e.g., the Palomar-Green (PG) Quasar sample; Schmidt \& Green 1983), quasars with similar optical properties exhibit very different properties in radio bands (Kellermann et al. 1989). The ratio of the radio flux at 5 GHz to the optical flux at $B$ band (i.e., the radio loudness parameter ${\cal R}\equiv f_{\rm 5~GHz}$/$f_{\rm B}$) has been adopted as an indicator of the radio properties of quasars (Kellermann et al. 1989). However, recent studies based on deep radio surveys, e.g., Faint Images of the Radio Sky at Twenty cm and NRAO VLA Sky Survey (Becker et al. 1995; Condon et al. 1998), and optical massive surveys, e.g., SDSS (Sloan Digital Sky Survey) and Two Degree Field Survey (York et al. 2000; Croom et al. 2001), suggest that the distribution of radio loudness of AGNs is not bimodal but rather continuous (e.g. Ivezic et al. 2002; Laor 2003). Understanding the distribution of radio loudness of AGNs is crucial for addressing the basic questions such as how jets are formed, accelerated and collimated, and why the efficiency of jet production can be so different among objects very similar in all other aspects, as well as the questions concerning jets in black hole and neutron star X-ray binaries (see review by Fender 2006).

Radio-quiet AGNs, for example the Seyferts, are usually hosted by late-type galaxies with ``under-luminous" or silent jets (e.g. Schawinski et al. 2011). On the other hand, it is known that radio-loud AGNs with strong radio jets almost never reside in the late type, i.e. spiral galaxies (e.g. Best et al. 2005, but also see Morganti et al. 2011). Blazars, including Flat-spectrum radio quasars (FSRQs) and BL Lacertae objects, are an extreme subclass of radio-loud AGNs. They are characterized by the luminous, rapidly variable, and polarized non-thermal continuum emissions, extending from radio to $\gamma$-ray (GeV and TeV) energies, which are widely accepted to be produced in the relativistic jets oriented close to the line of sight (Blandford \& Rees 1978; Ulrich et al. 1997). Their spectral energy distributions (SEDs) have a universal two-bump structure in log$\rm \nu F_{\nu}$-log$\rm \nu$ plot. The first bump is (likely) by synchrotron emission of relativistic electrons in magnetic fields while the second bump extending to $\gamma$ rays is usually explained as inverse Compton (IC) scattering of soft photons from either inside and/or outside of the jet by the same population of relativistic electrons (Maraschi et al. 1992; Dermer \& Schlickeiser 1993; Sikora et al. 1994; B{\l}a{\.z}ejowski et al. 2000). Besides of the typical Radio-loud quasars and Radio-quiet quasars, there are the so-called `radio-intermediate quasars' (RIQs, Falcke et al. 1996a,b; Wang et al., 2006). RIQs have compact radio cores at arcsecond scales with relatively high brightness temperatures and flat and variable radio spectra in common which are similar to radio-loud quasars. However, RIQs also possess low radio-to-optical ratio and weak extended steep-spectrum emission which are atypical of radio-loud quasars. Therefore, RIQs bridge the classical radio-loud and radio-quiet AGNs and can be used to probe the connection between these two canonical groups.

Since the CGRO era, it has been recognized that the blazars and radio galaxies are capable to generate strong $\gamma$-ray emissions (Hartman et al. 1999). In the current third {\it Fermi} Large Area Telescope (LAT, Atwood et al. 2009) source catalog (3FGL, Acero et al. 2015), the extragalactic $\gamma$-ray sky is dominated by radio-loud AGNs. The vast majority of these sources are blazars  (Ackermann et al. 2015; Liao et al. 2014). And $\gamma$-ray emissions from handful of radio-loud AGNs with misdirected jets (i.e., the so-called misaligned AGNs including radio galaxies and steep spectrum radio quasars) are also detected (Abdo et al. 2010a; Liao et al. 2015a). Except several nearby galaxies whose $\gamma$-ray emissions are mainly from the starburst activity (e.g. Abdo et al. 2010b; Hayashida et al. 2013), a new class of $\gamma$-ray AGNs, the radio-loud narrow-line Seyfert 1 galaxies, has been firmly established (Abdo et al. 2009a; Liao et al. 2015b). In this work we search for $\gamma$-ray emission from III Zw 2.

III Zw 2 is a triple galaxy group. The brightest source of the group, III Zw 2A (also named as PG 0007+106 or Mrk 1501 at the redshift of $z=0.089$, referred as III Zw 2 throughout this paper) is a AGN with Seyfert I nuclei (Zwicky 1967; Arp 1968), and also included in the PG quasar sample (Schmidt \& Green 1983). Its host galaxy was identified as a spiral (e.g. Hutchings \& Campbell 1983; Taylor et al. 1996). However, recent study of the bulge/disk decomposition on its $HST$ $H$-band image indicates an elliptical morphology (Veilleux et al. 2009). An extended low surface brightness emission among the galaxy group and a tidal bridge from III Zw 2 with several knots of star forming regions linking to a nearby companion III Zw 2B suggest an undergoing merging phase (Surace et al., 2001). Thus, considering the evidence of galaxy merging and following the morphological classification pattern of Schawinski et al. (2010), we refer the host galaxy of III Zw 2 as the indeterminate-type class. III Zw 2 is famous for its large-amplitude radio variability with over 20-fold increases at timescale of years (e.g. Aller et al. 1985). Correlated flux variations from radio to X-rays have been observed (Salvi et al. 2002). It is also the first Seyfert with detection of superluminal jet motion (the apparent jet speed lower limit is of 1.29$\pm$0.05 c, Brunthaler et al. 2000). Other evidence of hosting strong radio jet include a core-dominated flat-spectrum radio morphology with high brightness temperature core and hard X-ray spectrum (Falcke et al. 1996a; Chen et al. 2012). However, extended radio emission of III Zw 2 is rather weak compared with its core emission (Brunthaler et al. 2005; Cooper et al. 2007) and its radio-loudness is moderate (from several tens  to $\simeq$300, mainly due to the radio variability), which are typical behaviors of RIQs (Falcke et al. 1996a). Based on the violent radio variability and its multiwavelength spectral energy distribution (SED), $\gamma$-ray emission of III Zw 2 is expected (Chen et al. 2010; Lister et al. 2015). Searching for the $\gamma$-ray emission of III Zw 2 both in GeV and TeV energies in the past  however failed to yield significant detection (Aharonian et al. 2008; Abdo et al. 2009b; Teng et al. 2011; Ackermann et al. 2012).

In this paper, we carry out a detailed analysis of {\it Fermi}-LAT $\gamma$-ray data of III Zw 2, together with our optical multi-bands photometric data observed from Yunnan Observatories and other multiwavelength data from the public data archives and the literature. The paper is organized as follows: In Section 2 we describe the data analysis routines. The $\gamma$-ray emission characteristics of III Zw 2 are given in Section 3. Its multiwavelength variability properties are examined in Section 4. Finally, we present the discussion and summary of our study.

\section{OBSERVATION and DATA ANALYSIS}

\subsection{LAT Data Analysis}
The {\it Fermi}-LAT (Atwood et al. 2009) is a pair-conversion $\gamma$-ray telescope sensitive to photon energies greater than 20 MeV. The LAT has a large peak effective area ($\sim$8000 $\rm cm^{2}$ for 1 GeV photons), viewing $\simeq$ 2.4 sr of the full sky with angular resolution (68\% containment radius) better than $\rm 1^{\circ}$ at 1 GeV. In its routine survey mode, LAT performs a complete and uniform coverage of the sky in every 3 hours. Note that the recently released Pass 8 data set has been significantly improved in comparison with the former ones, including a wide energy reach (changing from the range of 0.1-300~GeV to 60~MeV-500~GeV), an enhanced effective area especially in the low energy range, and the better localization. All these lead to an improvement of 30\%-50\% enhanced differential point-source sensitivity (Atwood et al. 2013).

The {\it Pass} 8 data used in this paper were collected during the first 7-year operation of Fermi-LAT (i.e., from 2008 August 4th to 2015 August 4th). Photon events belonging to \texttt{evclass} 128 and \texttt{evtype} 3  with the energy ranging from 100 MeV to 500 GeV were taken into account. The updated standard {\it ScienceTools} software package version \texttt{v10r0p5} with the instrument response functions of \texttt{P8R2\_SOURCE\_V6} were adopted throughout the data analysis. For the LAT background files, we used \texttt{gll\_iem\_v06.fit} as the galactic diffuse model and \texttt{iso\_P8R2\_SOURCE\_V6\_v06.txt} for the isotropic diffuse emission template \footnote{http://fermi.gsfc.nasa.gov/ssc/data/access/lat/BackgroundModels.html}. The entire data set was filtered with \texttt{gtselect} and \texttt{gtmktime} tasks by following the standard analysis threads \footnote{http://fermi.gsfc.nasa.gov/ssc/data/analysis/scitools/}.

The \texttt{unbinned} likelihood algorithm (Mattox et al. 1996) implemented in the \texttt{gtlike} task was used to extract the flux and spectrum. All 3FGL sources within $20^{\circ}$ of the target were included. The flux and spectral parameters of sources within $10^{\circ}$ region of interest (ROI) together with normalization factors of the two diffuse backgrounds were set to be free, while parameters of other sources were fixed at the 3FGL values. For the sub-day light curve analysis, following the method adopted in Liao \& Bai (2015), we fixed the fluxes of the diffuse emission components at the values obtained by fitting the data collected over the flaring period. For the neighboring $\gamma$-ray emitters within the ROI at these special cases, their normalizations were still set free, while the spectral parameters were fixed as the 3FGL values. In the analysis we firstly added a presumed $\gamma$-ray source with power-law spectral template corresponding to III Zw 2 into the initial background model generated from \texttt{make3FGLxml.py}\footnote{http://fermi.gsfc.nasa.gov/ssc/data/analysis/user}. Its $\gamma$-ray position was initially set as the same as its radio position. After the model fit we checked the Test Statistic (TS) value of the target and made a scale residual TS map. If any $\gamma$-ray excess with TS value over 25 appeared in the TS map, we added a new source with a power-law spectrum into the background model to address it. The $\gamma$-ray locations of the target and the new background sources were obtained by task \texttt{gtfindsrc}. Then the updated background model was refitted to obtain the final result.

\subsection{Swift Data Analysis}
Since 2007, the space X-ray mission, {\it Swift} (Gehrels et al. 2004) has observed the source region 9 times. We excluded the observation taking on 2010 February 2 (Obs ID = 00036363004) because of too short exposure ($\sim 75$ s). When available, we analyzed both XRT and UVOT data with the FTOOLS software version 6.17. For XRT data, we performed the initial event cleaning with \texttt{xrtpipeline} using standard quality cuts, and then extracted the source spectra within circular regions with a radius of 20 pixels. The ancillary matrix files produced by the task \texttt{xrtmkarf} and the response matrix files (v012) were used for spectral analysis. We also grouped the spectra to have at least 20 counts per bin to ensure valid results using $\chi^{2}$ statistical analysis. By linking the value of the hydrogen column density, we fitted the left 8 observational data sets simultaneously by an absorbed power-law model and sumamrize the results in Table 1. The parameter of absorption (nH = $3.96_{-2.22}^{+2.23}$ $\times10^{20}$ cm$^{-2}$) is consistent with the values given by the data from other X-ray missions (e.g. Salvi et al. 2002). The UVOT has six filters: V, B, U, UVW1, UVM2, and UW2 with a coverage of 2.1$-$7.8 eV. We performed aperture photometry using \texttt{uvotsource} with a 5 arcsec circular aperture, and the background extraction was carried out in a larger source free region.

\subsection{Optical Observation and Data Analysis}
The variability of III Zw 2 was photometrically monitored in optical bands at Yunnan Observatories, making use of the 2.4 m telescope \footnote{http://www.gmg.org.cn} and the 1.02 m telescope \footnote{http://www1.ynao.ac.cn/$\sim$omt/}. Details about these telescopes have been introduced in Liao et al. (2014). The standard differential photometric procedure was followed. Sky flat field at dusk and dawn in good weather conditions and bias frames were taken at every observing night. Different exposure times were applied for various seeing and weather conditions. All frames were processed using bias and flat-field corrections by the task \texttt{CCDRED} package of the \texttt{IRAF} software, and the photometry was performed by the \texttt{APPHOT} package. Magnitudes of the source were calculated by with calibration stars in the image frame \footnote{https://www.lsw.uni-heidelberg.de/projects/extragalactic/charts/}. Observing uncertainty of every night was the root-mean-square (RMS) error of differential magnitude between two calibration stars. At least one of them must be fainter than or as bright as the source (Bai et al. 1998, 1999). The multi-band magnitudes observed from Yunnan Observatories are listed in Table 2.

Complementary optical data were derived from the Catalina Real-time Transient Survey (CRTS, Drake et al. 2009; Djorgovski et al. 2011). The photometry was transformed from the unfiltered instrumental magnitude to Cousins $V$  by $V$  = $V_{CSS}$ + $0.31(B-V)^{2}$ + 0.04 \footnote{http://nesssi.cacr.caltech.edu/DataRelease/FAQ2.html\#improve}. We averaged the values obtained during the same observing night.  The CRTS daily light curve starts from MJD 53706 to MJD 56223 with 94 data points, nearly one observation per month, performing a well coverage at timescale of years.  The correction for the interstellar extinction and the color excess of the observed optical/UV magnitudes were adopted (Schlafly \& Finkbeiner 2011; Cardelli et al. 1989). And optical photometric data were converted from magnitude system to flux in Jansky (Bessell 2005).  Consistence between different optical observation systems was checked. Note that the CRTS fluxes tend to be systematically higher than fluxes from others. For example, an UVOT observation at MJD 55929.5 gave a $V$ band flux density of 2.1 mJy while a CRTS observation at MJD 55930.1 provided a $V$ band flux density of 4.1 mJy. Since III Zw 2 was then at optical low state, such a discrepancy could not be caused by intrinsic variability. Thus, the CRTS light curve was only used for the purpose of exhibition of long timescale variability trend.

\subsection{Radio Data}
III Zw 2 is included in the Owens Valley Radio Observatory (OVRO) 40 m telescope monitoring program\footnote{http://www.astro.caltech.edu/ovroblazars/}. This program encompasses over 1500 objects above declination of $20^{\circ}$, most of which are blazars, with observations for each source twice per week at a frequency of 15 GHz (Richards et al. 2011). III Zw 2 is also included in the Monitoring of Jets in Active Galactic Nuclei With VLBA Experiments (MOJAVE\footnote{http://www.physics.purdue.edu/astro/MOJAVE/index.html}) program (Lister et al. 2009), from which multi-epoch VLBA observations at 15~GHz of several hundreds of the brightest, most compact radio sources in the northern sky have been carried out. Recently, radio data of III Zw 2 from the MOJAVE program have been published, including core and jet light curves as well as jet motion features (Lister et al. 2016).

\section{RESULTS}
\subsection{Detecting significant $\gamma$-ray emission from III Zw 2}
We initially perform a fit to the entire 7-year {\it Fermi}-LAT data and do not find a significant $\gamma$-ray source in the direction of III Zw 2. For the \texttt{unbinned} likelihood analysis, the TS value of the presumed $\gamma$-ray source is only $\simeq$15 ($<4\sigma$), as shown in Figure 1, consistent with its absence in any $\gamma$-ray source catalogs. For the tentative gamma-ray emission, a photon flux of (6.6$\pm$2.1)$\times 10^{-9}$ ph $\rm cm^{-2}$ $\rm s^{-1}$ and a photon index of 3.0$\pm$0.3 are obtained (we refer to a spectral index $\alpha$ as the energy index such that  $F_{\nu}\propto\nu^{-\alpha}$, corresponding to a photon index $\Gamma_{\rm ph} = \alpha+1$). And the corresponding isotropic $\gamma$-ray luminosity in the range from 100~MeV to 500~GeV is (4.6$\pm$1.3)$\times 10^{43}$ erg $\rm s^{-1}$ (throughout this paper we adopt a $\Lambda$CDM cosmology model with $\rm H_{0}$ = 67 km $\rm s^{-1} Mpc^{-1}$, $\Omega_{m}$ = 0.32, $\Omega_{\Lambda}$ = 0.68, Planck Collaboration et al. 2014).

We then search for possible short-term $\gamma$-ray outburst. Firstly, the year-bin $\gamma$-ray light curve is extracted (see Figure 2). Except for the second and sixth bins, the TS values of the emission in other time bins are below 1. The TS value of the emission in the sixth time bin is also relatively low, which is $\simeq$6 (i.e., $<3\sigma$). However, the TS value of the $\gamma$-ray emission in the second time bin is as high as 38 ($>5\sigma$). We also extract monthly $\gamma$-ray light curve for the whole 7 years LAT data. Except for the second year LAT data, the highest TS value of the monthly bin is about 5, so III Zw 2 has not been detected during most of the LAT observational time. The rather weak signal appears in the sixth year LAT data is likely due to the background fluctuation. On the other hands, two $\gamma$-ray flares have been identified in the monthly $\gamma$-ray light curve in the second year Fermi-LAT data. Then further the 2 days time bin $\gamma$-ray light curves are extracted during these two periods. The start and end of the two flaring epochs are selected as from MJD 55120 to MJD 55190, and MJD 55322 to MJD 55368, respectively (see Figure 3 and 4). Individual \texttt{gtlike} analyses for these two epochs give the TS values of 35.3 and 50.9, respectively. Such significant signals are confirmed by the TS maps, see Figure 1. Since III Zw 2 is a high Galactic latitude source ($|l|>50^{\circ}$), contamination from uncertainty of Galactic diffuse emission is negligible. And during such short time periods, the newly emerging $\gamma$-ray source is one of the most dominate source within the ROI. The only source with comparable $\gamma$-ray photon flux is $>7^{\circ}$ away. We hence conclude that the detection of $\gamma$-ray emission is robust. Furthermore, localization of the central excess is performed. The $\gamma$-ray position of R.A. $2.621^{\circ}$ and decl. $11.1232^{\circ}$ is obtained for the flare in 2009, and R.A. $2.440^{\circ}$ and decl. $10.9349^{\circ}$ is for the other flare in 2010, with corresponding 95\% Confidential Level (C.L.) error radius of $810^{\prime\prime}$ and $781^{\prime\prime}$, respectively. Considering that the angular separations between the radio position and $\gamma$-ray locations are $535^{\prime\prime}$ and $683^{\prime\prime}$, respectively, in both cases III Zw 2 falls into the 95\% C.L. $\gamma$-ray location error radius, as shown in Figure 5. We also seek other potential counterparts through the SIMBAD database\footnote{http://simbad.u-strasbg.fr/simbad/}, especially for radio-loud AGNs.  III Zw 2 is found to be the only radio-loud AGN within the 95\% C.L. $\gamma$-ray location error radius. With these well-established facts we conclude that these significant $\gamma$-ray signals are from III Zw 2.

Besides of the detailed individual $\gamma$-ray analyses for the two flaring epochs, we also perform a joint analysis to increase the significance of the $\gamma$-ray detection and the TS value increase to 83. The $\gamma$-ray localization has got improved (the location error is $658^{\prime\prime}$), which is still larger than the source separation of $470^{\prime\prime}$, rendering III Zw 2 as the only radio-loud AGN candidate within the 95\% C.L. $\gamma$-ray location error radius (see Figure 5). The updated $\gamma$-ray location is R.A. $2.523^{\circ}$ and decl. $11.0536^{\circ}$, and all $\gamma$-ray results throughout this paper are based on this position. A joint analysis is also performed by adopting the \texttt{Pass 7 REP} data, similar results are obtained but with a much lower TS value $\simeq$ 38.

Single power-law function provides an acceptable description of the $\gamma$-ray spectrum of III Zw 2 during the flaring state, i.e.,
\begin{equation}
 \frac{dN}{dE}=(5.25\pm0.78)\times10^{-12}(\frac{E}{\rm 374.1~MeV})^{-(2.53\pm0.14)},
\end{equation}
and the photon flux is (9.7$\pm$1.7)$\times 10^{-8}$ ph $\rm cm^{-2}$ $\rm s^{-1}$. Note that it is over one order of magnitude than the 7-year averaged flux. The isotropic $\gamma$-ray luminosity is (9.6$\pm$1.6)$\times 10^{44}$ erg $\rm s^{-1}$. No significant improvement of the fit is found when more sophisticated spectral models are used. We also perform individual spectral analysis for each flares and the spectral indexes do not change significantly.

It is worth noting that signs of intraday $\gamma$-ray variability are shown in the 2-day light curves. For the flare in 2009, a fast flux decline is detected,  from (3.6$\pm$1.3)$\times 10^{-7}$ ph $\rm cm^{-2}$ $\rm s^{-1}$ at MJD 55140.1 to (0.8$\pm$0.7)$\times 10^{-7}$ ph $\rm cm^{-2}$ $\rm s^{-1}$ at MJD 55142.1. Adopting the classic method, $\tau_{\rm d}=\triangle t\times \ln2$/ln($F_{1}$/$F_{2}$), the flux doubling timescale is estimated as $\simeq$ 0.9 day. However, the relatively low TS values prevent a further investigation. On the other hand,  intraday $\gamma$-ray light curves are extracted for the flare in 2010. From the 12-hour light curve, it is remarkably to see that III Zw 2 is undergoing a giant flare during 12 hours, see Figure 3. This flare suddenly appeared with a peak flux of (1.2$\pm$0.3)$\times 10^{-6}$ ph $\rm cm^{-2}$ $\rm s^{-1}$ and the TS value is high up to $\sim 73$. Since 3-hour bin is the smallest bin allowed by the standard {\it Fermi}-LAT data analysis procedure with the standard software, the $\gamma$-ray light curve in such a short time bin is extracted to derive the minimum $\gamma$-ray variability timescale of III Zw 2, also see Figure 3. $\gamma$-ray flux maintains as a high value, $\gtrsim$ $10^{-6}$ ph $\rm cm^{-2}$ $\rm s^{-1}$ within 12 hours, and 3-hour peak flux is (1.7$\pm$0.6)$\times 10^{-6}$ ph $\rm cm^{-2}$ $\rm s^{-1}$. It is roughly 250 times of the 7-year average $\gamma$-ray flux. Such a large-amplitude vaiability is extreme even for blazars. In the ascent phase, $\gamma$-ray flux raises from $(0.3\pm0.2) \times 10^{-6}$ to $(1.5\pm0.6) \times 10^{-6}$ ph $\rm cm^{-2}$ $\rm s^{-1}$ within 6 hours. A corresponding flux doubling timescale is estimated as $\simeq 2.6$ hr. A simple exponential function can well describe the 3-hr light curve,
\begin{equation}
F(t) = 2F_{0}[(e^{(t_{0}-t)/T_{\rm var}}+e^{(t-t_{0})/T_{\rm var}}]^{-1},
\end{equation}
where $F_{0}$ and $t_{0}$ are set as flux and time of the peak, respectively. The variability time scale is estimated as about 3.4 hr and the corresponding doubling timescale is about 2.4 hr. Since the light curve is extracted during the survey mode operation of LAT, due to the limited exposure time, doubling timescales of III Zw 2 should be treated as upper limits only. Despite of the extreme large variability amplitude, such a short doubling timescale ($\sim 2-3$ hours) are detected at only a handful blazars (e.g. Foschini et al. 2011; Liao \& Bai 2015).

The rapid $\gamma$-ray variation allows us to make a constraint on the Doppler factor, avoiding a heavy absorption from the soft photons within the radiation radius through the $\gamma\gamma$ process (e.g. Begelman et al. 2008). The synchrotron emission is considered as the target photons. The optical depth of $\gamma\gamma$ absorption between the $\gamma$ rays and the soft photons can be simply calculated as (Dondi \& Ghisellini 1995):
\begin{equation}
\tau_{\gamma\gamma}(x^{\prime})=\frac{\sigma_{\rm T}}{5}n^{\prime}(x^{\prime}_{\rm t})x^{\prime}_{\rm t}R^{\prime},
\end{equation}
where $\sigma_{\rm T}$ is the scattering Thomson cross section, $n^{\prime}(x^{\prime})$ is the number density of the target photon, $x^{\prime}_{\rm t}$ is the energy of the target photon in dimensionless units, and $R^{\prime}$ is the absorption length. The doubling time of 2.5 hr is used to constrain the emission region. The absorption length is radius of the emission blob, $R^{\prime}\leq ct_{\rm var}\delta(1+z)^{-1}$. And luminosity of the absorbing X-ray synchrotron emission at several keVs is set as a relatively low value,  $\rm 10^{43}$ erg $\rm s^{-1}$ (Salvi et al. 2002). A lower limit of Doppler factor is then constrained as $\delta\gtrsim$ 9. For the $\gamma$-ray flare in 2009, a similar constraint can be also estimated, i.e., $\delta\gtrsim 6$.

\subsection{Multi-wavelength variability of III Zw 2}
\subsubsection{Simultaneous optical and $\gamma$-ray variability in Nov. 2009}
In the classic leptonic radiation model of FSRQs, both the non-thermal optical and $\gamma$-ray emissions are generated from the same population of high energy electrons and hence simultaneous variability of these emissions is expected. Actually, for $\gamma$-ray FSRQs, such predictions have been observed (e.g. Bonning et al. 2012). And the simultaneous optical and $\gamma$-ray variation is a powerful tool to identify the association between the $\gamma$-ray source and the counterpart. We have monitored III Zw 2 in Oct. and Nov. 2009 at Yunnan Observatories. Fluxes in three bands (i.e., Johnson $R$, $B$ and Cousins $I$) all exhibit brightening around Nov. 2009 (see Figure 2). A zoomed-in of the flaring epoch is shown in Figure 4. The $R$ band light curve is adopted to study optical variability during this epoch. When $R$ band magnitudes are unavailable, simultaneous $I$ band magnitudes are extrapolated into $R-$band based on the observed $R-I$ color. Within 15 days, the optical flux of III Zw 2 exhibited an increase of about 40\%. The peak time of the optical flare was at MJD 55145.6. If the optical monitoring time was over one hour, we also searched for optical intraday variability. However, no significant intraday variability is found. The peak flux is about 3 times as the flux at the optical low status, e.g. MJD 55832.8. Such a large optical variability amplitude is likely from the jet of III Zw 2. The 2009 $\gamma$-ray flare peaked on MJD 55140, suggesting that the optical and $\gamma$-ray flares were simultaneous (i.e., the optical flare lagged behind the $\gamma$-ray flare less than one week, consistent with studies of other FSRQs;  Bonning et al. 2012). Because the detection of the 2009 $\gamma$-ray flare alone is significant ($>5\sigma$), together with the simultaneous optical and $\gamma$-ray flares, we claim the discovery of $\gamma$-ray emission of III Zw 2. Furthermore, the optical spectral variability is studied by the simultaneous $R$ and $B$ photometric observations. No significant spectral variability is found during the flaring epoch.  And also no significant spectral difference is detected comparing to the SDSS observation at optical low state (see Figure 6), indicating that the accretion disk emission significantly contribute to the optical/UV domain even for the case of 2009 $\gamma$-ray flare.

\subsubsection{ X-ray and radio variability}
Because the {\it Swift} observations were sparse and there were no X-ray observations during two $\gamma$-ray flares, it is impossible to directly study the connection between $\gamma$-ray and X-ray emissions. However, we note that X-ray flux at MJD 55385, just 45 days after the $\gamma$-ray flare at MJD 55340, is roughly two times of other fluxes, see Table 1. Nevertheless, comparing to the historical X-ray data with 10-fold X-ray variability and 1-2 keV X-ray flux as high as $\simeq 10^{-11}$ erg $\rm cm^{-2}$ $\rm s^{-1}$ (Salvi et al. 2002),  the X-ray variability observed by {\it Swift} is not extreme. Because of low signal-to-noise ratio of the {\it Swift} data, single power-law function is adopted to describe the X-ray spectra and more complicated features reported in the literature can not be checked (e.g. Piconcelli et al. 2005). The X-ray photon indices maintain as $\simeq$ 1.7, consistent with other studies (e.g. Salvi et al. 2002).

Strong radio variability is observed by OVRO, as shown in Figure 2. The maximum and minimum fluxes of the 15~GHz OVRO light curve are about 1.82 and 0.08~Jy respectively, indicating an over 20-fold variability, consistent with the literature (e.g. Aller et al. 1985). Different from rapid optical and $\gamma$-ray variability usually lasting for only several days, the typical timescale of radio flare is as long as several hundred days. There are four main flares in the OVRO light curve and none of these radio flares simultaneously coincided with $\gamma$-ray flares. Cross-correlation analyses between $\gamma$-ray and radio light curves for {\it Fermi} bright blazars suggest that the radio emission typically lags the $\gamma$-ray emission for a few months (e.g. Fuhrmann et al. 2014; Max-Moerbeck et al. 2014). Interestingly, the peaking time of the strongest radio flare in the OVRO light curve is at MJD 55223, about 80 days behind the first $\gamma$-ray flare at MJD 55140. Moreover, in the descent phase of this radio flare, there is a plateau indicative a radio sub-flare at MJD 55432, which is also about 80 days behind the the second $\gamma$-ray flare at MJD 55340. However, there is no $\gamma$-ray flares corresponding to other radio flares. Since the radio core and the parsec jet are resolved by the MOJAVE VLBA observations, radio light curves and angular offset for these two components are provided (Lister et al. 2016), see Figure 2. The core light curve is well consistent with the OVRO light curve because III Zw 2 exhibits a core-dominated radio morphology. It is interesting that the pc jet flux at MJD 55389 (0.13 Jy) is significantly higher than it at MJD 54985 (0.007 Jy), together with a shorter offset from the core, see Figure 2. These evidence suggest there is a new ejecta coming out since MJD 55389. And pc jets from MJD 53046 to MJD 54985 and from MJD 55389 to 56445 are marked as different jet components, with apparent speeds of $\beta_{\rm app}=1.2\pm0.07$ and  $\beta_{\rm app}=1.58\pm0.29$, respectively (Lister et al. 2016). Note that the observation of the newly emerging ejecta is just about 50 days after the $\gamma$-ray flare at MJD 55340 when the violent intraday $\gamma$-ray variability is detected, indicating that the $\gamma$-ray flare may link to the ejection of the new jet knot, see the MOJAVE radio image\footnote{http://www.physics.purdue.edu/astro/MOJAVE/sourcepages/0007+106.shtml} (Figure 7). Such ejection speeds are significantly lower than that of other $\gamma$-ray FSRQs ($\beta_{\rm app} >$ 10) with detection of $\gamma$-ray variability at timescale of hours (e.g. Jorstad et al. 2005). Dense VLBA observations coincident with future $\gamma$-ray flaring epoch might yield a $\beta_{\rm app}$ much higher than that inferred from the long-term observational data.

\section{DISCUSSIONS}
Since the last piece information of electromagnetic emission (i.e. the $\gamma$ rays) of III Zw 2 has been collected, its radio to gamma-ray SED has been investigated for the first time. A homogeneous one-zone synchrotron plus IC model is used to calculate the jet emission. The broadband electromagnetic emission comes from a compact homogeneous blob with relativistic speed with a
radius of $R$ embedded in the magnetic field. A broken power-law spectrum for particle distribution has been assumed, i.e.,
\begin{equation}
N(\gamma )=\left\{ \begin{array}{ll}
                    K\gamma ^{-p_1}  &  \mbox{ $\gamma_{\rm min}\leq \gamma \leq \gamma_{br}$} \\
            K\gamma _{\rm br}^{p_2-p_1} \gamma ^{-p_2}  &  \mbox{ $\gamma _{\rm br}<\gamma\leq\gamma_{\rm max}$,}
           \end{array}
       \right.
\end{equation}
The model parameters include $R$, the magnetic field strength $B$, electron break energy $\gamma_{br}$, the minimum and maximum energies, $\gamma_{min}$ and $\gamma_{max}$, of the radiating electrons, the normalization of the particle number density $K$, and the indices $p_{1,2}$ of the broken power-law particle distribution. The synchrotron self-absorption and the Klein-Nishina effect in the IC scattering are properly addressed in our calculations. $\chi^{2}$-minimization method is used to obtain the best-fitting input parameters. Specially, the $B$ and $\delta$ are constrained at 1$\sigma$ confidence level, based on the fit probability $p\propto e^{-\chi^{2}/2}$ where $\chi^{2}$ values are calculated for wide ranges of $B$ and $\delta$. A similar SED modeling strategy has been adopted in previous studies (e.g. Zhang et al. 2009; Liao et al. 2014).

Simultaneous {\it Planck}, {\it Swift}, and {\it Fermi} observations for III Zw 2 at the beginning of July in 2010 provide the best coverage of its electromagnetic emission (Giommi et al. 2012). In addition, {\it WISE} perform a simultaneous complementary infrared observation at MJD 55381.6 (Wright et al. 2010). Moreover, infrared colors are given as $w1-w2=0.93$ and $w2-w3 =2.53$, suggesting that III Zw 2 falls into the {\it WISE} blazar stripe and the jet emission is significant in infrared bands (Massaro et al. 2012). Together with the coinstantaneous $V$ band UVOT observation, the peak frequency of the synchrotron bump can be constrained in optical/near-infrared range (see Figure 8). However, despite of the well sampled data from radio to X rays, there is no significant simultaneous $\gamma$-ray detection. Based on our $\gamma$-ray temporal analysis, this multiwavelength campaign is performed several tens of days after the violent variability in MJD 55340. Due to its nature of rapid variability, it is reasonable to assume that III Zw 2 was then under the detection threshold of {\it Fermi}-LAT. Nevertheless, the 7-year average $\gamma$-ray spectrum is believed as a good approximate and used instead of the simultaneous $\gamma$-ray upperlimit. Note that in our modeling several data points have been excluded. The UV emission is likely dominated by the accretion disk emission. The W4 band {\it WISE} infrared flux is clearly deviated from the smooth nonthermal emission. For the {\it Planck} radio data, due to the significant spectral break around 200~GHz, only two data at the 353 and 545~GHz are included. On the other hand, thanks for the optical observations by Yunnan Observatories, SED of III Zw 2 including simultaneous optical and $\gamma$-ray detections is obtained at the first time. However, coverage of the jet emission is rather limited. The OVRO 15~GHz and Yunnan Observatories $B$ band data are not considered in the SED modeling. Due to lacking of simultaneous sub-mm/infrared and X-ray data, the locations of the synchrotron and IC peaks are unknown, which makes the SED input parameters highly unconstrained in this case. Nevertheless, because the $\gamma$-ray doubling time of 0.9 day is detected then, the Doppler factor is set as 6, and the radius of the emission blob can be constrained as $\rm 1.3\times10^{16}$ cm, $\rm R\leq ct_{\rm var}\delta(1+z)^{-1}$. Other parameters including $p_{1}$ and $\gamma_{min}$ are set as same as the other case. For both SED modelings, the soft photons of external Compton (EC) process are assumed to be from the broad emission lines, especially the Lyman $\alpha$ line. And the energy density of the soft photons is set as $\rm 3.84\times10^{-4}$ erg $\rm cm^{-3}$, consistent with the literature (Chen et al. 2012).

The simple leptonic model can provide acceptable descriptions of the SEDs from both epochs (see Figure 8). And the input parameters are listed in Table 3. Assuming that one proton corresponds to one relativistic emitting electron and that protons are `cold' in the comoving frame (Celotti \& Ghisellini 2008), the jet power for the SED with $\gamma$-ray detection in 2009 is estimated as $\sim 8\times10^{44}$ erg $\rm s^{-1}$, and for the other case in 2010 the instantaneous jet power is $\sim 4\times10^{44}$ erg $\rm s^{-1}$, suggesting that the jet power of III Zw 2 is comparable with other $\gamma$-ray FSRQs (Ghisellini et al. 2010; Zhang et al. 2015). However, unlike other $\gamma$-ray FSRQs whose $\gamma$-ray emissions are dominated by the EC process and hence the Compton Dominances are far above 1, for III Zw 2, the EC component is not significant even in the $\gamma$-ray flaring state and the IC component is as luminous as the synchrotron one. One possible explanation is that III Zw 2 possesses a lower energy density of the BLR emission ($\rm \simeq 4\times10^{-4}$ erg $\rm cm^{-3}$) in comparison with a typical value $\rm \simeq 3\times10^{-2}$ erg $\rm cm^{-3}$ (Ghisellini et al. 2012). The two sets of input parameters are compared to investigate the reason of brightening of $\gamma$-ray emission. The increase of the Doppler factor may play an important role for the $\gamma$-ray flare.

Strong $\gamma$-ray emission as well as large amplitude and rapid gamma-ray variability, the distinct characters of blazars, are also detected for III Zw 2. And SED modeling of III Zw 2 suggests that it possesses a strong jet similar with $\gamma$-ray FSRQs. It is somehow `unusual' that a RIQ can possess such a blazar behavior. We then seek differences between $\gamma$-ray FSRQs and III Zw 2. We note that radio loudness is actually not appropriate for comparison. It is influenced by the violent radio variability and possible contamination from the host galaxy. The extended radio luminosity is then adopted since it is not significantly variable and not suffered from the Doppler beaming effect. Recently, 1.4~GHz VLA extended radio luminosites and intrinsic bolometric luminosities of 128 2FGL FSRQs (Nolan et al. 2012) have been provided in Nemmen et al. (2012). For III Zw 2, its 1.4~GHz VLA observation gives the extended radio flux density of 17~mJy (Cooper et al. 2007), with corresponding luminosity of 5.3$\times10^{39}$ erg $\rm s^{-1}$. And the intrinsic bolometric luminosity of III Zw 2, derived from the isotropic bolometric luminosity by the modeling of the SED in 2010, is $L_{\rm bol}=(1-\cos(1/\Gamma))L_{\rm bol}^{\rm iso} \simeq 3\times10^{44}$ erg $\rm s^{-1}$, where $\Gamma=3$ is the Lorenz factor based on our SED modeling study. Then III Zw 2 is plotted into the $L_{\rm bol}$-$L_{\rm radio}^{\rm ext}$ diagram, see Figure 9. Interestingly, although III Zw 2 shares similar $\rm L_{bol}$ with those FSRQs, the $L_{\rm radio}^{\rm ext}$ of III Zw 2 is generally lower than the FSRQs. Radio spectral and spatial evolution study of III Zw 2 indicates that it shares the same physical processes with young radio sources (Brunthaler et al. 2005). And in addition of its weak radio lobe emission reaching up to 20-30 kpc (e.g. Falcke et al. 1999; Brunthaler et al. 2005), III Zw 2 could be the case of a recurrent core jet activity in conjunction with a relic radio lobe. Since the extended radio emission is widely accepted from the accumulated old-age low energy electrons, so it is reasonable that $L_{\rm radio}^{\rm ext}$ of III Zw 2 is generally lower in comparison with the FSRQs due to its recently active core.

Finally, let us jump out from the frame of the radio loud AGN alone. Recently, accumulated evidence suggest galaxies convolved with its central SMBHs (e.g. Kormendy \& Ho 2013; Heckman \& Best 2014). Radio jets likely play an unique role of the AGN feedback (e.g. Fabian 2012). Galaxy Merger is believed as a possible path for generating the radio loud AGNs (e.g. Hopkins et al. 2008; Lagos et al. 2009). In one way, Galaxy Merger may be to lower the specific angular momentum of gas in the galaxy and thus to drive the gas toward to the center, and another effect of mergers is  spin-up and to increase the mass of the SMBH (Chiaberge et al. 2015). The spin of BH is believed as a crucial factor for generating strong radio jet, the so-called spin paradigm (Blandford et al. 1990). For III Zw 2, tentative evidence of broad Fe $\rm K_{\alpha}$ emission line may be indicative of fast spin of its central BH (Jim{\'e}nez-Bail{\'o}n et al. 2005), agreeing with the existence of strong jet based on it strong gamma-ray emission and rapid and large-amplitude gamma-ray variability. Besides of these parameters, the magnetic flux threading the SMBH is also proposed as a dominant factor in launching powerful jets (Sikora \& Begelman 2013). And the radio activity of RIQs is associated to the fluctuating magnetic field of jets, such as the magnetic reconnection. Interestingly, rapid $\gamma$-ray variability with doubling time of nearly 2 hours detected for III Zw 2 could respond to such a physical process. Moreover, III Zw 2 is hosted by a merging galaxy and considering the evidences of rejuvenation radio core mentioned in former paragraph, it is likely to be a ideal target which provides insights of the fuelling and triggering of the activity in radio loud AGNs. Further simultaneous multiwavelength observations is urgently needed to study III Zw 2.

In summary, we present the results of the radio to $\gamma$-ray observations of III Zw 2 and report the discovery of its $\gamma$-ray emission. Although significant $\gamma$-ray emission is not detected from the 7-year averaged {\it Fermi}-LAT data, III Zw 2 exhibits strong $\gamma$-ray emission in short term. The TS values of $\gamma$-ray signals in Nov. 2009 and May 2010 are all above 25 and the joint analysis gives a TS value as high as 83. Results of $\gamma$-ray localization and coincided $\gamma$-ray and optical variations in Nov. 2009 strongly support the association between III Zw 2 and the $\gamma$-ray source. Moreover,  violent $\gamma$-ray variability with doubling timescale of 2.5 hours is detected in another $\gamma$-ray flare in 2010 when the 3-hour $\gamma$-ray flux is 250 times of the 7-year average flux. Such an extreme variability behavior indicates that the central engine which generates the moderate-relativistic powerful jet is likely similar to those of blazars. This scenario is also supported by modeling the simultaneous spectral energy distributions of III Zw 2. On the other hand, considering its radio core similar with young radio sources and relic radio lobe extending to 20-30 kpc, III Zw 2 likely harbors a recurrent activity core and serves as a valuable target for investigating the fuelling and triggering of the activity in radio loud AGNs.

\acknowledgements
We appreciate the helpful suggestions from the anonymous referee, which led to a substantial improvement of this work. We acknowledge the support of the staff of the Lijiang 2.4 m and Kunming 1 m telescopes. Fund for these telescopes has been provided by CAS and the People's Government of Yunnan Province. This research has made use of data obtained from the High Energy Astrophysics Science Archive Research Center (HEASARC), provided by $\rm NASA^{\prime}$s Goddard Space Flight Center. This research has also made use of the NASA/IPAC Extragalactic Database and the NASA/IPAC Infrared Science Archive which are operated by the Jet Propulsion Laboratory, California Institute of Technology, under contract with the National Aeronautics and Space Administration. This research makes use of the SIMBAD database, operated at CDS, Strasbourg, France. This research has made use of data from the OVRO 40 M Telescope Fermi Blazar Monitoring Program which is supported by NASA under awards NNX08AW31G and NNX11A043G, and by the NSF under awards AST-0808050 and AST-1109911. This research has made use of data from the MOJAVE database that is maintained by the MOJAVE team (Lister et al., 2009, AJ, 137, 3718). The CSS survey is funded by the National Aeronautics and Space Administration under grant no. NNG05GF22G issued through the Science Mission Directorate Near-Earth Objects Observations Program. The CRTS survey is supported by the US National Science Foundation under grants AST-0909182and AST-1313422.

This work was supported in part by the National Basic Research Program of China (No. 2013CB837000), NSFC under grants 11525313 (i.e., Funds for Distinguished Young Scholars), 11361140349, 11433009, 11133006,  11233006, 11303022, 11673013 and 11361140347, as well as Joint Research Fund in Astronomy (U1431123). This work was also supported by the Strategic Priority Research Program ``The emergence of Cosmological Structures" of the Chinese Academy of Sciences (grant No.XDB09000000), and the Key Research Program of the Chinese Academy of Sciences (grant No. KJZD-EW-M06).

\begin{figure}
\centering

\includegraphics[scale=0.36]{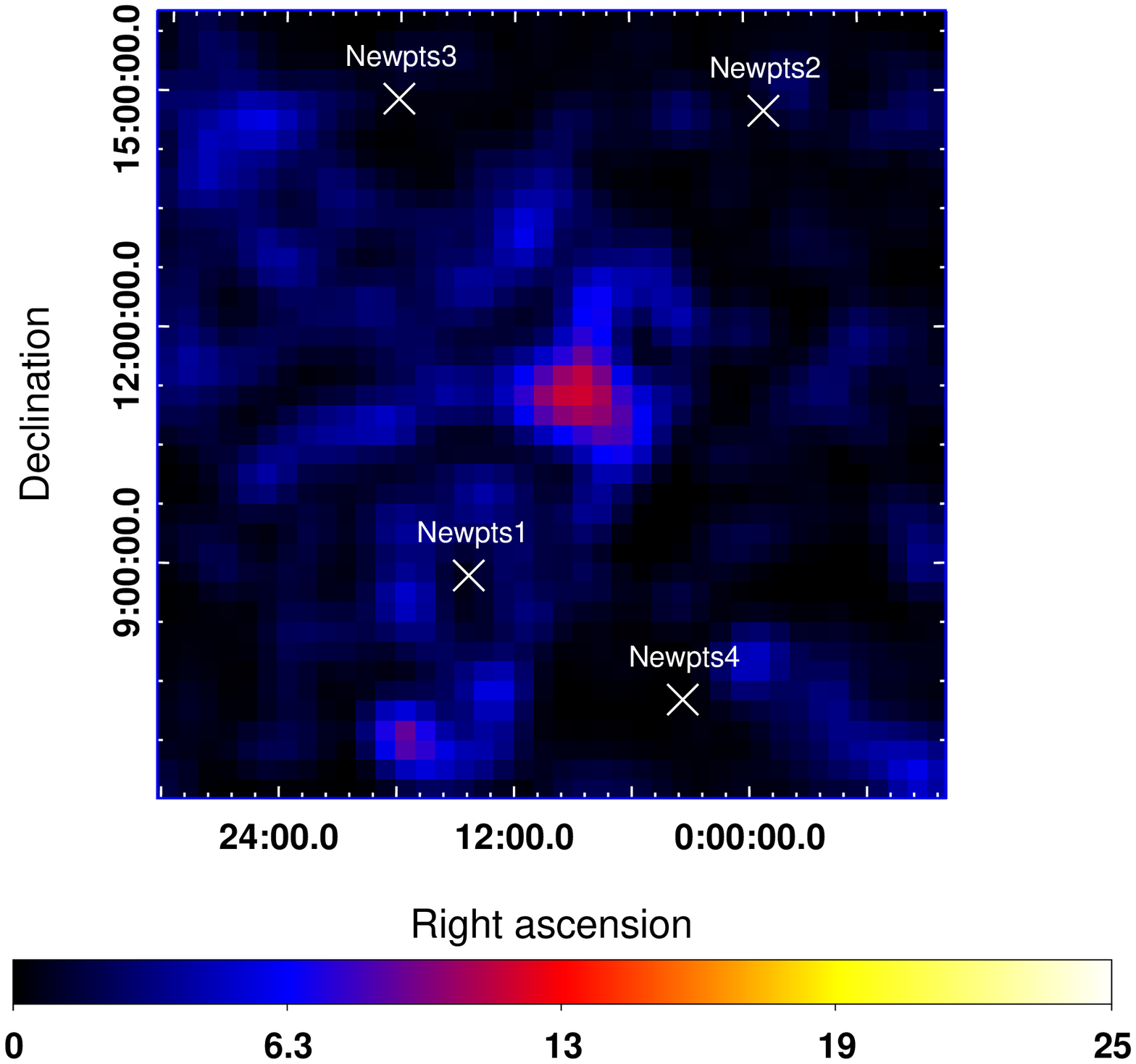}
\includegraphics[scale=0.36]{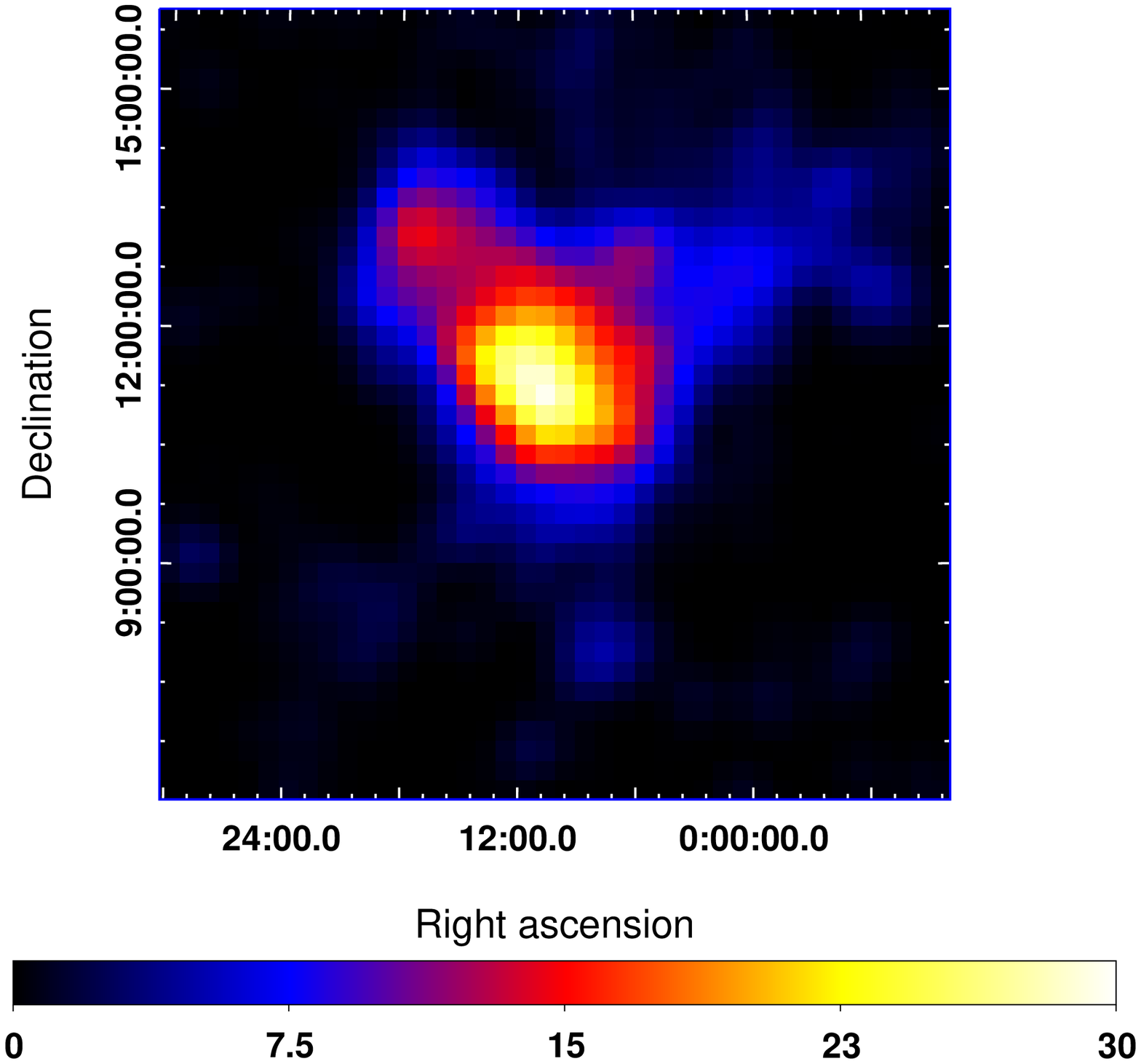}
\includegraphics[scale=0.36]{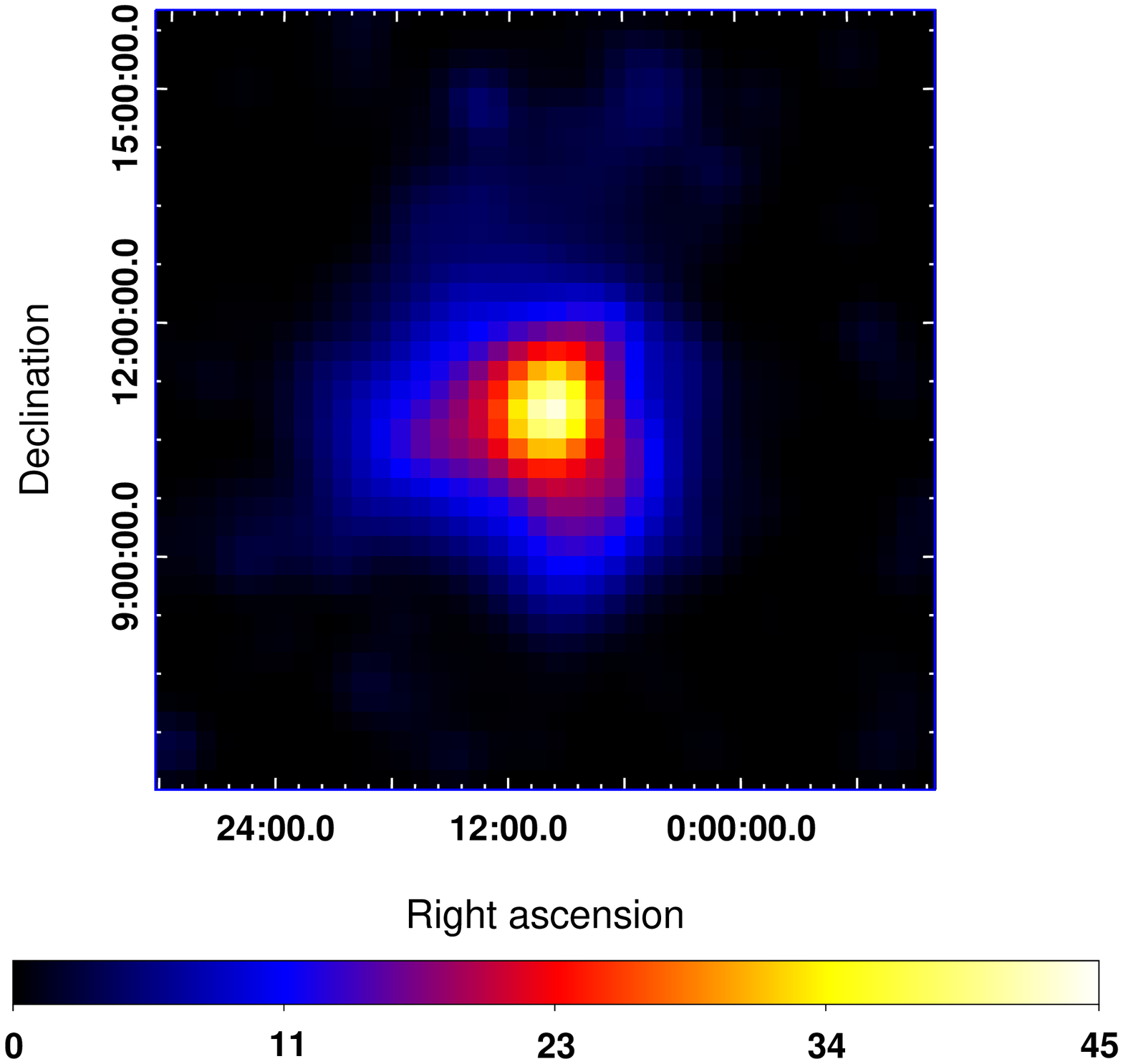}
\includegraphics[scale=0.36]{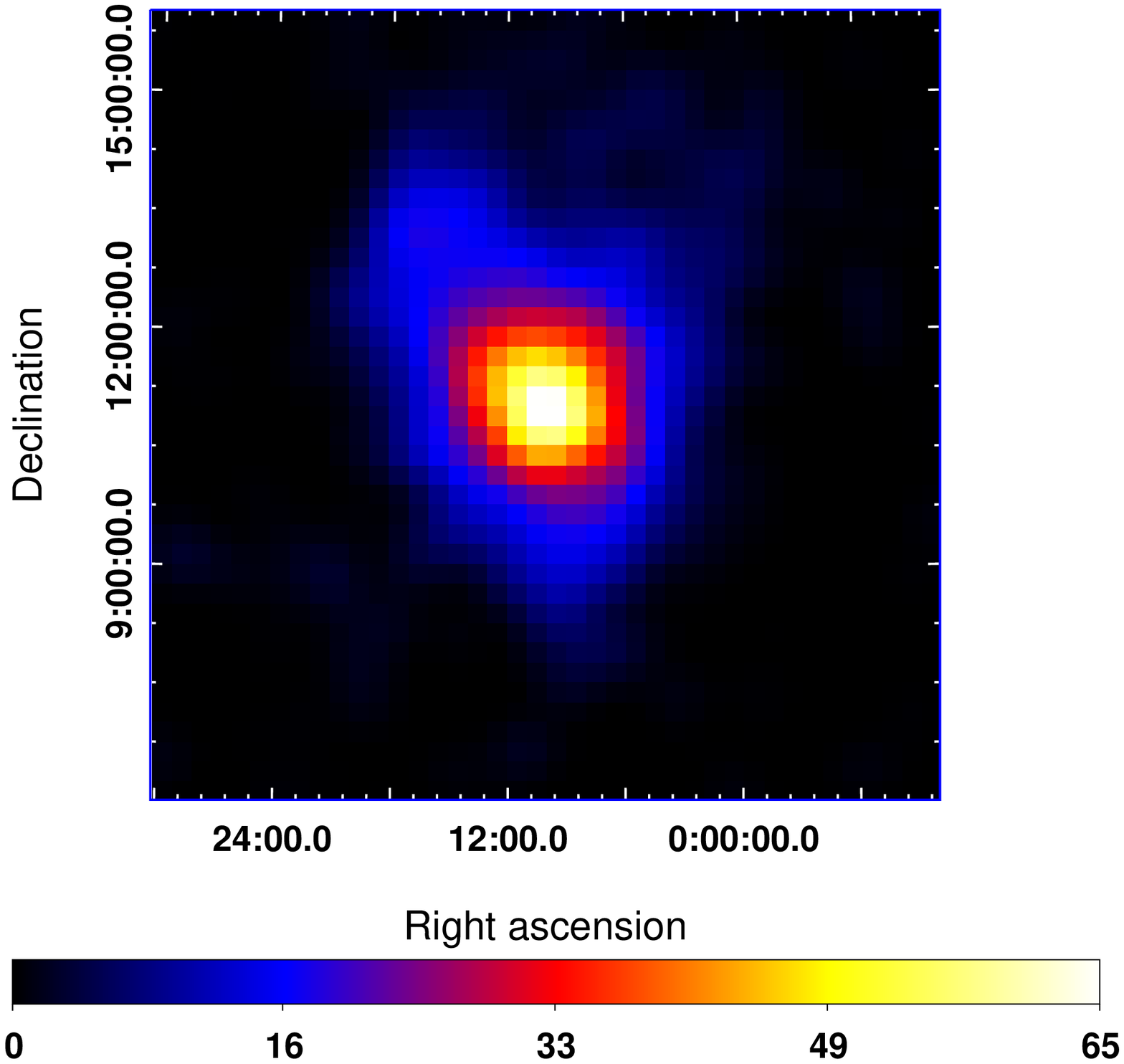}

\caption{TS map of photons from 100 MeV to 500 GeV for $10^{\circ} \times 10^{\circ}$ region centered at III Zw 2. Upper left panel is for the entire 7-year LAT data; upper right panel represents to $\gamma$-ray flare in Nov. 2009 and bottom left is for another flare in May 2010; bottom right is for the joint analysis. The diffuse backgrounds, 3FGL and additional sources are subtracted. TS values of the central excess corresponding to III Zw 2 are consistent with {\it gtlike} analyses. The newly emerged $\gamma$-ray neighbors within $5^{\circ}$ in the 7-year TS map are marked in white color. The map is smoothed with $\sigma$=$0.2^{\circ}$ Gaussian function.}
\label{Fig.1}
\end{figure}

\begin{figure}
\centering
\includegraphics[scale=0.7]{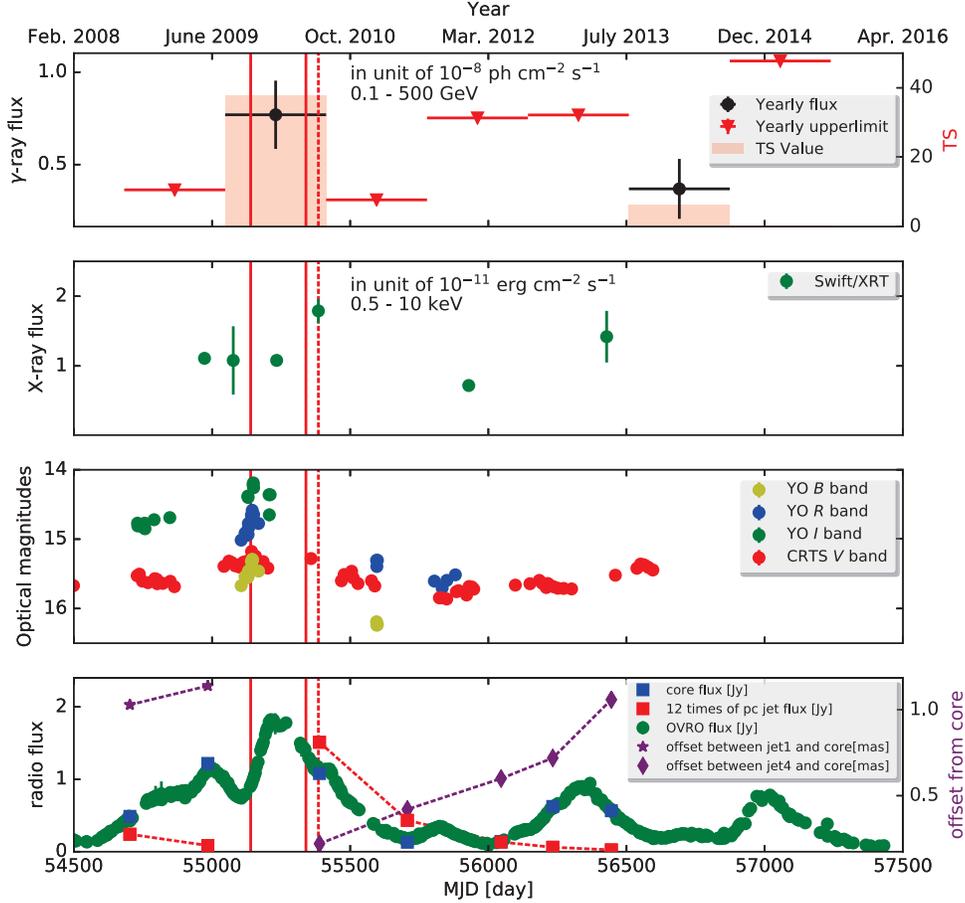}
\caption{The multiwavelength light curves of III Zw 2. In the $\gamma$-ray panel, the $\gamma$-ray fluxes are in unit of $\rm 10^{-8}$ ph $\rm cm^{-2}$ s$^{-1}$, the black circles correspond to the fluxes and the red triangles are the 2$\sigma$ upper limits. The red bars are the TS values in each data bin. In the X-ray panel, unabsorbed 0.5$-$10 keV X-ray fluxes are in units of 10$^{-11}$  erg cm$^{-2}$ s$^{-1}$. In the optical panel, green, blue and yellow circles are the magnitudes of $I$, $R$ and $B$ bands observed in Yunnan Observatories (YO), respectively. The red circle are CRTS $V$ band magnitudes minus 0.8 mag. In the radio panel, the green circles are OVRO single-dish fluxes, the blue and red squares are the MOJAVE core and parsec jet VLBA fluxes. The purple stars and diamonds respond to offset between core and pc-jet for two different jet. The two solid red vertical lines mark the peaking times of the two $\gamma$-ray flare and the dashed red vertical line respond to the time of the simultaneous {\it Planck}, {\it Swift}, and {\it Fermi} campaign.}
\label{Fig.2}
\end{figure}

\begin{figure}
\centering
\includegraphics[scale=0.8]{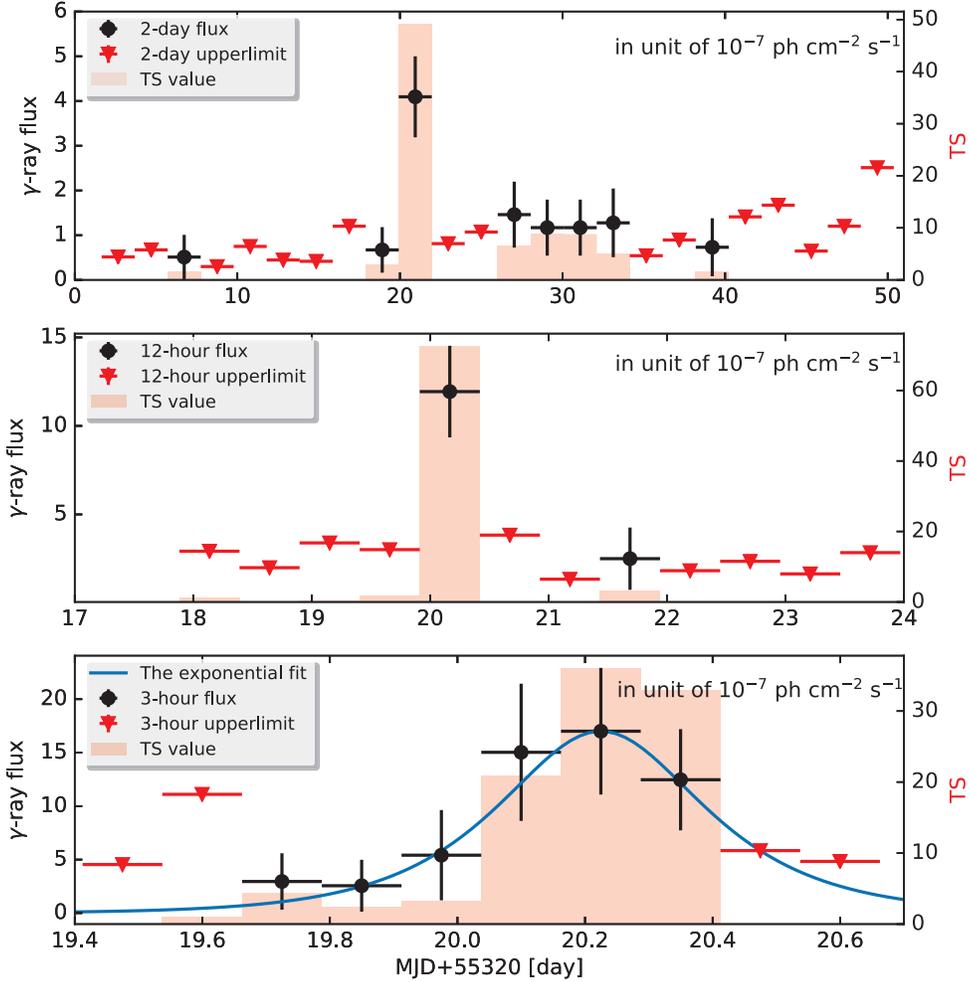}
\caption{The ``zoomed-in" $\gamma$-ray light curve for the $\gamma$-ray flare in May 2010. The $\gamma$-ray fluxes are in unit of $\rm 10^{-7}$ ph $\rm cm^{-2}$ s$^{-1}$, the black circles represent the fluxes and the red triangles are the 2$\sigma$ upper limits. The red bars are the TS values in each data bin.}
\label{Fig.3}
\end{figure}

\begin{figure}
\centering
\includegraphics[scale=0.8]{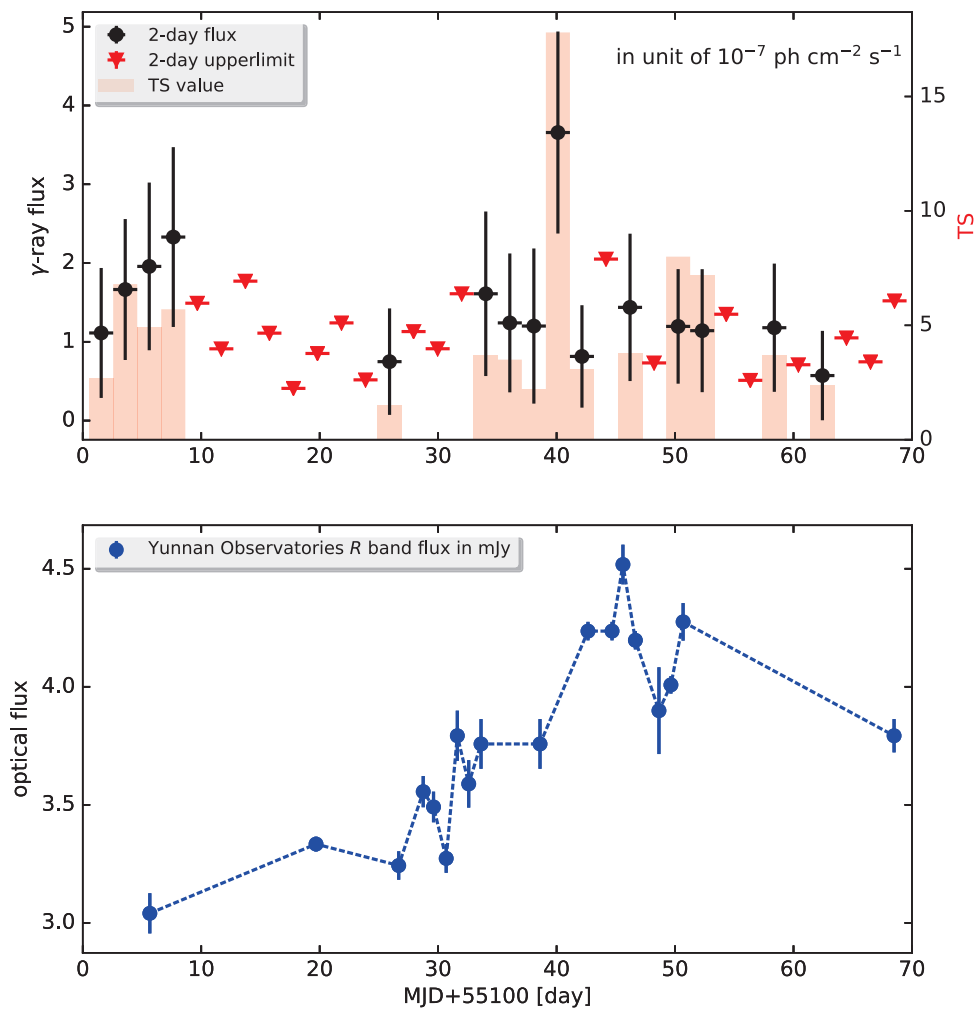}
\caption{Simultaneous $\gamma$-ray and optical flares in Nov. 2009. In the $\gamma$-ray panel, the $\gamma$-ray fluxes are in unit of $\rm 10^{-7}$ ph $\rm cm^{-2}$ s$^{-1}$, the black circles represent the fluxes and the red triangles are the 2$\sigma$ upper limits. The red bars are the TS values in each data bin. In the optical panel, the blue circle are $R$ band fluxes including also that extrapolated from the $I$ band magnitudes by adopting the $R-I=0.6$ mag.}
\label{Fig.4}
\end{figure}

\begin{figure}
\centering
\includegraphics[scale=0.6]{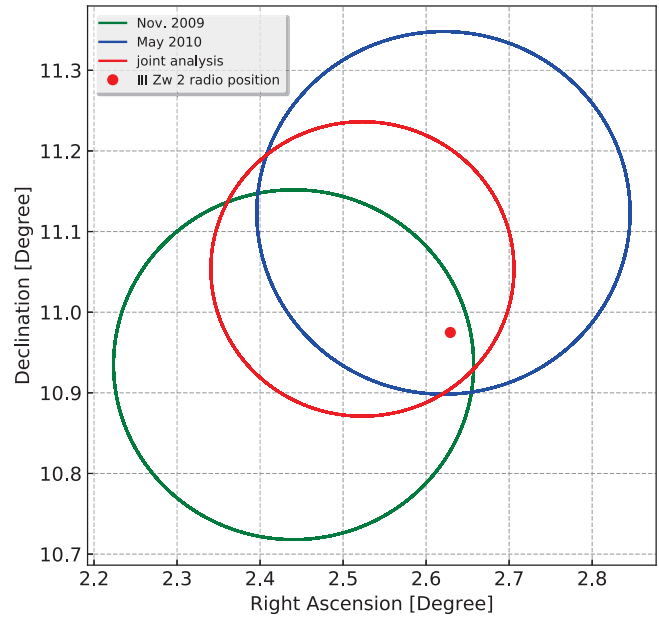}
\caption{Results of $\gamma$-ray localization. The blue, green and red circle lines correspond to the 95\% C.L. $\gamma$-ray location radii for the flare epochs in Nov. 2009 and May 2010, as well as the joint analysis, respectively. The red dot marks the radio position of III Zw 2.}
\label{Fig.5}
\end{figure}

\begin{figure}
\centering
\includegraphics[scale=0.6]{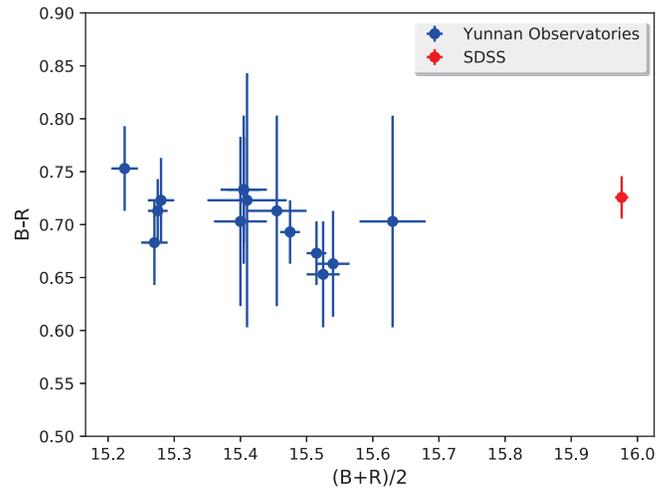}
\caption{Optical spectral variability of III Zw 2. The blue circles are optical colors from simultaneous optical monitoring by Yunnan Observatories in Oct. and Nov. of 2009. The red circle is from SDSS observation at MJD 54771 when III Zw 2 is at a relatively optical low state.}
\label{Fig.6}
\end{figure}

\begin{figure}
\centering
\includegraphics[scale=0.6]{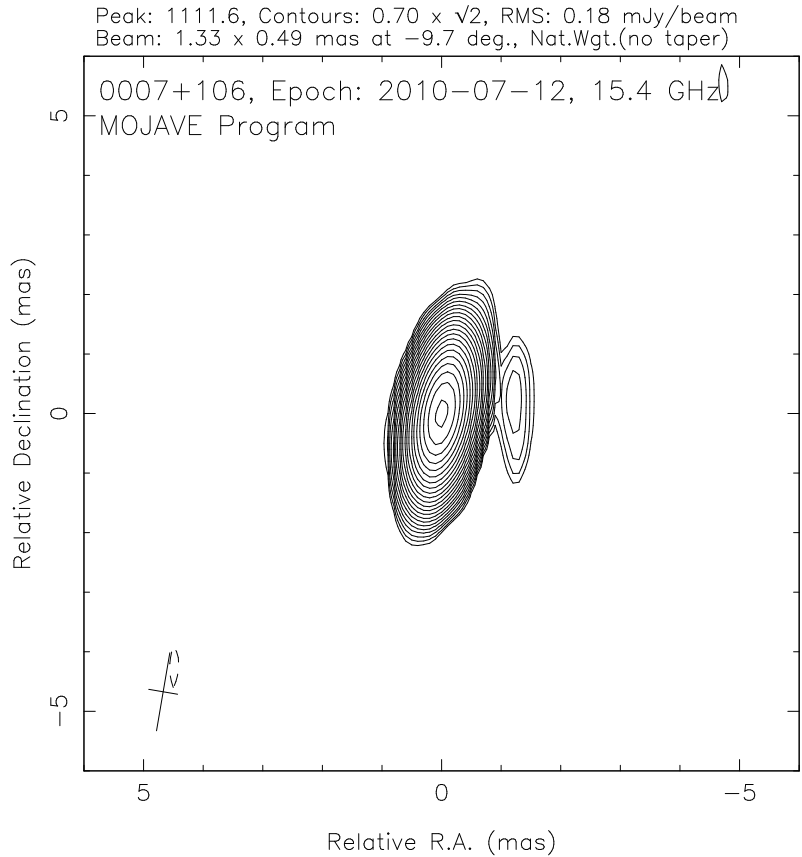}
\caption{The 15~GHz VLBA radio image of III Zw 2 directly derived from the MOJAVE data website, where an obvious parsec jet can be seen. At this time, the parsec jet flux significantly brightens and a new component begins to emerge (see also Figure 1). And it is only 50 days after the violent $\gamma$-ray variability in 2010, suggesting that the $\gamma$-ray flare may link to a new ejecta.}
\label{Fig.7}
\end{figure}

\begin{figure}
\centering
\includegraphics[scale=0.8]{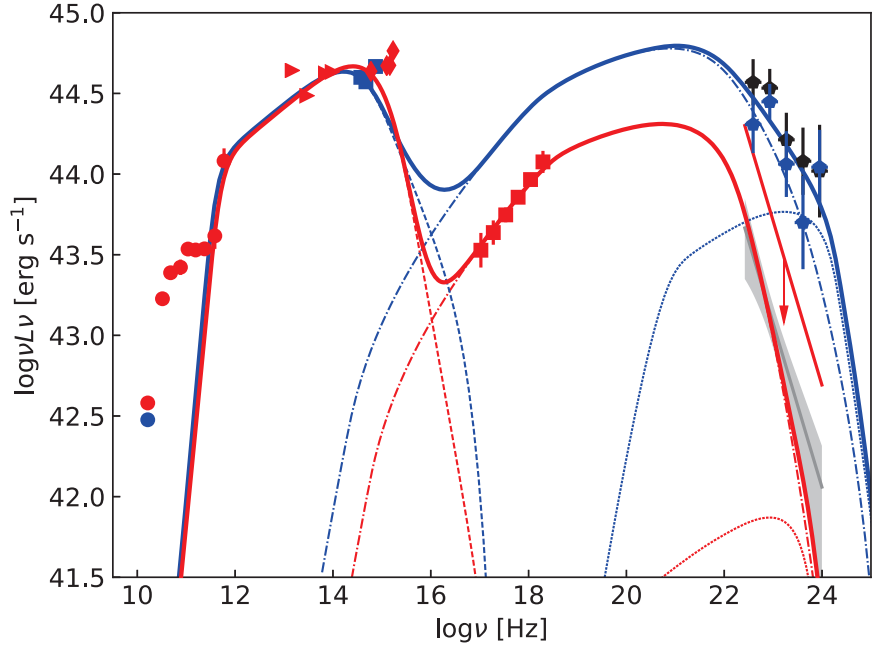}
\caption{SEDs and the jet emission modelings. The blue and red colors respond to SEDs and data points with and without $\gamma$-ray detection, respectively. Blue data points include: OVRO data (circle), Yunnan Observatories data (squares) and LAT data in 2009 (pentagrams). Red data points include: {\it Planck} data (circles), {\it WISE} data (triangles), {\it Swift}/UVOT data (diamonds), {\it Swift}/XRT (squares) and simultaneous LAT upper limit (red line with a downward arrow). In addition, the black line is the best fit of entire 7-year {\it Fermi}-LAT data and the grey butterfly is its 1$\sigma$ uncertainty area. Finally, the black pentagrams are LAT data for the 2010 $\gamma$-ray flare. For the calculated jet emission, the solid, dashed, dashed dotted and the dotted lines correspond to the total, synchrotron, SSC and EC components, respectively.}
\label{Fig.8}
\end{figure}

\begin{figure}
\centering
\includegraphics[scale=0.6]{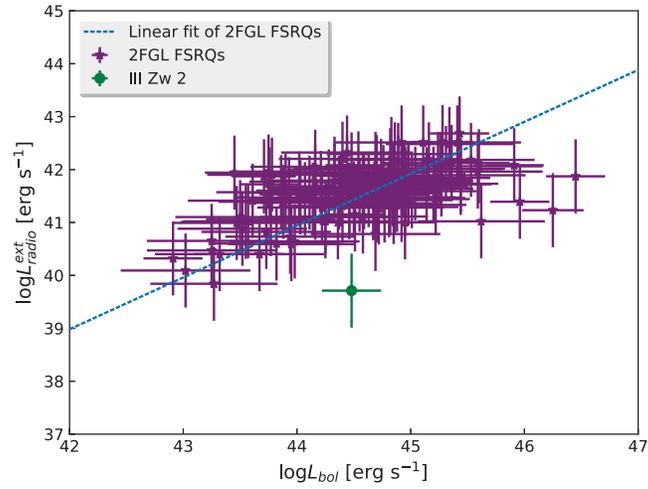}
\caption{The position of III Zw 2 in the blazar $L_{\rm bol}$ - $L_{\rm radio}^{\rm ext}$ diagram. The 2FGL FSRQ data, marked as purple stars, are adopted from Nemmen et al. (2012). The blue dashed line is the best linear fit for 2FGL blazars (adopted from Nemmen et al. 2012). III Zw 2 is marked as a green circle. The uncertainties of $L_{\rm bol}$ and  $L_{\rm radio}^{\rm ext}$ of III Zw 2 are adopted as 0.25 and 0.7 dex, i.e., the sample average values found in Nemmen et al. (2012).}
\label{Fig.9}
\end{figure}

\clearpage

\begin{deluxetable}{lcccccccccccc}
\tabletypesize{\small} \tablenum{1} \tablewidth{0pt} \tablecaption{{\it Swift} results} \tablehead{\colhead{ObsID} & \colhead{Date}  &
\colhead{$\Gamma_{X}$} & \colhead{$F_{X}$} & \colhead{V} & \colhead{UW1} & \colhead{UM2} & \colhead{UW2}} \startdata
\hline

00036363001  & 2007-06-21  & $1.60_{-0.10}^{+0.11}$ & $1.11_{-0.08}^{+0.08}$ & ... & ... & ... & ... \\
00036363002  & 2009-05-22  & $1.66_{-0.10}^{+0.10}$ & $1.11_{-0.08}^{+0.08}$ & ... & ... & ... & ...  \\
00036363003  & 2009-09-03  & $1.98_{-0.69}^{+0.84}$ & $1.08_{-0.27}^{+0.49}$ & ... & ... & ... & ...  \\
00036363005  & 2010-02-07  & $1.69_{-0.09}^{+0.10}$ & $1.08_{-0.07}^{+0.07}$  & ... & ... & ... & ... \\
00036363006  & 2010-07-08  & $1.55_{-0.14}^{+0.15}$ & $1.80_{-0.21}^{+0.21}$  & $3.68\pm{0.09}$ & $2.09\pm{0.05}$ & $1.62\pm{0.03}$ & $1.65\pm{0.03}$ \\
00036363007  & 2010-07-08  & $1.64_{-0.11}^{+0.12}$ & $1.77_{-0.14}^{+0.14}$  & ... & ... & ... & $1.63\pm{0.03}$ \\
00036363008  & 2012-01-03  & $1.64_{-0.14}^{+0.14}$ & $0.72_{-0.08}^{+0.08}$  & $2.07\pm{0.06}$ & $0.73\pm{0.02}$ & $0.57\pm{0.03}$ & $0.54\pm{0.01}$ \\
00049402001  & 2013-05-16  & $1.43_{-0.31}^{+0.33}$ & $1.42_{-0.34}^{+0.39}$  & ... & $1.21\pm{0.02}$ & $0.94\pm{0.02}$ & $0.95\pm{0.02}$
\enddata

\tablecomments{The absorption column density for estimation of X-ray flux is constrained as $3.96_{-2.22}^{+2.23}\times 10^{20}$ cm$^{-2}$, with $\chi^2$/dof of 211.0/216.
 Unabsorbed 0.5$-$10 keV X-ray fluxes are in units of 10$^{-11}$  erg cm$^{-2}$ s$^{-1}$. The extinction magnitudes in four UVOT bands are calculated as: $A_{V}=0.275$, $A_{UW1}=0.582$, $A_{UM2}=0.820$, $A_{UW2}=0.718$. The absorption corrected optical/UV fluxes are in unit of mJy.
 \label{swift}}
\end{deluxetable}

\begin{deluxetable}{lrrrrrr}
\tablenum{2} \tablewidth{0pt}
\tablecaption{The multi-bands photometric data from Yunnan Observatories\tablenotemark{1}}
\tablehead{ \colhead{MJD\tablenotemark{2}} &\colhead{Mag.\tablenotemark{3}} &\colhead{SigMag.\tablenotemark{4}} &\colhead{Band\tablenotemark{5}} }

\startdata
54373.71 &14.58  &0.09 &I  \\[3pt]
54400.63 &14.70  &0.03 &I  \\[3pt]
\enddata
\tablenotetext{1}{Table 1 is available in its entirety in machine-readable forms in the online journal.
A portion is shown here for guidance regarding its form and content.}
\tablenotetext{2}{The observation date}
\tablenotetext{3}{The nightly average magnitude. The correction for the interstellar extinction has been already completed}
\tablenotetext{4}{Uncertainty of magnitude}
\tablenotetext{5}{The photometric band }
\end{deluxetable}

\begin{deluxetable}{lccccccccccc}
\scriptsize
\tablenum{3} \tablewidth{0pt}
\tablecaption{Input parameters of the SED models\tablenotemark{a}}
\tablehead{ \colhead{Epoch} &\colhead{$\rm p_{1}$} &\colhead{$\rm p_{2}$} &\colhead{$\gamma_{br}$\tablenotemark{b}} &\colhead{$\gamma_{min}$} &\colhead{K ($\rm cm^{-3}$)} &\colhead{B (Gauss)} &\colhead{$\delta$} &\colhead{R (cm)}
}

\startdata
Nov. 2009 &2.5 &4.5 &$\rm 3.6\times10^{3}$ &40 &$\rm 1.6\times10^{5}$ &1.0\tablenotemark{c} &6 &$\rm 1.3\times10^{16}$ \\[3pt]
July 2010 &2.5 &6.4 &$\rm 3.2\times10^{3}$ &40 &$\rm 4.9\times10^{5}$ &5.6$\pm$0.6 &3.0$\pm$0.1 &$\rm 1.7\times10^{16}$ \\[3pt]
\enddata

\tablenotetext{a}{A detailed description of the parameters is provided in the Discussion Section.}

\tablenotetext{b}{$\gamma_{max}$ is set as a hundred times of $\gamma_{br}$.}
\tablenotetext{c}{In this case, the SED data is so limited that further constraints of the uncertainties of B and $\delta$ can not be given.}

\end{deluxetable}

\end{document}